\documentstyle[12pt]{article}

\newcommand{\ab}{{(a)}}
\newcommand{\bb}{{(b)}}
\newcommand{\brho}{\bar \rho}
\newcommand{\btau}{\bar \tau}
\newcommand{\bt}{\bar t}
\newcommand{\br}{\bar r}

\newcommand{\be}{\begin{equation}}
\newcommand{\ee}{\end{equation}}
\newcommand{\ben}{\begin{eqnarray}\displaystyle}
\newcommand{\een}{\end{eqnarray}}
\newcommand{\refb}[1]{(\ref{#1})}
\newcommand{\p}{\partial}
\newcommand{\sectiono}[1]{\section{#1}\setcounter{equation}{0}}

\title{EXTREMAL BLACK HOLES AND ELEMENTARY STRING STATES}

\author{Ashoke Sen \\
\\
International Centre for Theoretical Physics \\
P.O. Box 586, I-34100 Trieste, ITALY \\
and \\
Tata Institute of Fundamental Research \\
Homi Bhabha Road, Bombay 400005, INDIA \\
sen@theory.tifr.res.in, sen@tifrvax.bitnet
}

\begin{document}

\maketitle

\begin{abstract}

Some of the extremal black hole solutions in string theory have
the same quantum numbers
as the Bogomol'nyi saturated elementary string states. We explore the
possibility that these black holes can be identified to
elementary string excitations. It is shown that stringy effects could
correct the Bekenstein-Hawking formula for the black hole entropy
in such a way that it correctly reproduces the
logarithm of the density of elementary string states.
In particular, this entropy
has the correct dependence on three independent parameters,
the mass and the left-handed charge of the black hole,
and the string coupling constant.

\end{abstract}

\vfill

\vbox{ \hbox{hep-th/9504147} \hbox{TIFR-TH-95-19} \hbox{April, 1995}}
\hfill ~

\eject


There have been suggestions from diverse points of view that black holes
should be treated as elementary
particles\cite{HAWS,HOOFT,HOLW,SUSS,RUSSO,DUFF,SBLACK,HULL,WITTEN,STROM,NEW}.
The purpose of
this paper is to investigate one aspect of this suggestion for a specific
class of black holes in supersymmetric string theories\cite{REVIEWS},
known as extremal black holes saturating the Bogomol'nyi
bound\cite{BOGOM}.

Treating black holes as elementary particles poses a
puzzle. In string theory there is an infinite tower of massive states
which have the same quantum numbers as classical electrically charged
black hole solutions. The question that arises naturally is, should
we count the black holes and elementary string states as separate
elementary particles? Or do they correspond to different ways of
representing the same states? The second alternative certainly looks
much more attractive. But to further substantiate this claim, one must
show that the black holes have the same properties as elementary
string states besides carrying the same quantum numbers.

One of the features which seems to be common between black holes and
elementary string states is that for both the degeneracy
of states with given mass (and charge) increases very rapidly with mass.
For elementary string states this growth is due to the large number of
oscillator states that correspond to a state with a given mass.
For black holes, it arises due to the fact that the classical
Bekenstein-Hawking entropy, which is proportional to the area of the
event horizon, increases very rapidly with mass. Unfortunately,
in trying to push this analogy further, we run into trouble.
For elementary string states, the logarithm of the
degeneracy of states increases linearly with mass, whereas the
Bekenstein-Hawking entropy of the black hole increases as the square of
the mass.

It has been suggested\cite{SUSS,RUSSO} that this difficulty might be
circumvented by taking into account the large renormalization of the mass
of a black hole. In
particular, it has been argued that black holes of mass $M^2$ should be
identified to elementary string excitations of mass $M$. This would
remove the discrepancy between the two entropies.

There are, however,
some particular states in string theory, which do not receive any
mass renormalization\cite{WITOLIVE,KALLOSH}.
These are the Bogomol'nyi saturated states alluded to
earlier, and in comparing them to (extremal) black holes, we can no longer
appeal to any possible renormalization of the mass. Thus these states
provide  a suitable laboratory for testing the hypothesis of black holes -
elementary string excitations correspondence. It is for this reason that
we shall focus our attention on these states in this paper. (This point has
also been advocated by Vafa\cite{VAFA}.)

How then are the entropies of extremal black holes related to the
degeneracies of Bogomol'nyi saturated elementary string states?
While the logarithm of the degeneracy of
Bogomol'nyi saturated elementary string states continues to grow
linearly with mass (although with a different proportionality factor),
the area of the event horizon of an extremal black hole of the kind we
are discussing actually vanishes. This is fortunate, since if it had not
vanished, it would have almost certainly been proportional to the square
of the black hole mass, leading to a contradiction, since now we can no
longer appeal to a mass renormalization.\footnote{It has however been
argued\cite{HAW} that the original calculation of ref.\cite{GIBHAW} has
to be modified for calculating the entropy of an extremal black hole. The
net result is that the entropy of an extremal black hole always
vanishes, irrespective of whether the area of the event horizon vanishes
or not. Since in the present case the area of the event horizon does
vanish, we do not need to appeal to the arguments of ref.\cite{HAW}.}
But now we have the opposite problem. The degeneracy associated with
extremal black hole states seems to be smaller than the degeneracy of
the elementary string states with the same quantum numbers.

To resolve
this puzzle, we could postulate that the entropy of the extremal black hole
is not exactly equal to the area of the event horizon, but the area
of a surface close to the event horizon, which we shall call the `stretched
horizon'. This assumption is not totally
unreasonable, since for various reasons we expect that our standard
understanding of the physics of the black hole has to be modified very
close to the horizon / singularity; and the stretched horizon
represents a surface
beyond which our standard understanding breaks down.
The relevant question then is, what
determines the location of the stretched horizon? Once we find an answer
to this question, we can compute its area and study its relationship
with the degeneracy of elementary string states. In particular, we can ask
if it has the correct dependence on the mass and charge of the black hole,
and the string coupling constant. The important point
to note is that in order to carry out this comparison,
we must define the location
of the stretched horizon in a way that does not invole any unknown function
of these three parameters. As we shall see, this can indeed be done.

We can explore different alternatives. If we believe that the necessary
modification of the physics near the horizon comes from the string
world-sheet effects, then we should define the stretched horizon
to be the surface where the string world sheet theory becomes strongly
coupled. In other words, this is the surface where the space-time
curvature associated with the string metric, and/or other target space
field strengths, becomes large. We shall see that
this definition determines the location of the stretched horizon
uniquely up to a purely numerical factor.

We might also try to
define the stretched horizon to be the place where the
string coupling constant becomes large. However, for black holes
of the type we shall discuss, this does not happen. In fact, the string
coupling vanishes as we approach the event horizon. Thus if we start
from a configuration where the asymptotic value of the string coupling
is small, then it remains small everywhere in space-time, and as a result
we cannot define the stretched
horizon using the criteria of string coupling becoming strong.

The third possibility makes use of black hole thermodynamics, and is
in fact the original definition of stretched horizon given in
ref.\cite{SUSS,SUSS2}. According to this definition,
the stretched horizon is defined to be the surface where the
local Unruh temperature for an observer, who is stationary in the
Schwarzschild coordinate, is of the order of
the Hagedorn temperature of the string
theory.\footnote{It has now been demonstrated\cite{DABH} that
the conventional
treatment of black hole thermodynamics does break down beyond this
surface.} This surface is close to the event horizon, where the local Unruh
temperature is infinite.
As we shall see, this definition actually coincides with the first
definition of the stretched horizon for electrically charged extremal black
holes.

In the rest of this paper, we shall carry out a careful calculation of the
area of the stretched horizon, defined as above, as a function of the mass
and charge of the black hole, as well as the string coupling constant, and
show that the result agrees with the logarithm of the degeneracy of the
elementary string states.\footnote{A preliminary version
of this result was stated briefly in ref.\cite{SBLACK}.}
We shall also discuss how stringy effects might produce such a
modification of the Bekenstein-Hawking formula.
Throughout this paper we shall work with the four dimensional theory
obtained by toroidal compactification of the heterotic string
theory\cite{NARAIN}, and use the normalization conventions of
ref.\cite{SREV}.

We begin by writing down the effective action describing the dynamics
of the massless fields in four dimensions:
\ben \label{en1}
S & = & {1\over 32 \pi} \int d^4 x \sqrt{-G} e^{-\Phi} \Big[
R_G + G^{\mu\nu} \p_\mu \Phi \p_\nu \Phi - {1\over 12}
G^{\mu\mu'} G^{\nu\nu'} G^{\rho\rho'} H_{\mu\nu\rho} H_{\mu'\nu'\rho'}
\nonumber \\
&& - G^{\mu\mu'} G^{\nu\nu'} F^\ab_{\mu\nu} (LML)_{ab} F^\bb_{\mu'\nu'}
+ {1\over 8} G^{\mu\nu} Tr(\p_\mu M L \p_\nu M L) \Big]\, .
\een
Here $G_{\mu\nu}$, $B_{\mu\nu}$ and $A^\ab_\mu$ ($0\le \mu,\nu\le 3$,
$1\le a \le 28$) are the string metric, anti-symmetric tensor fields,
and $U(1)^{28}$ gauge fields respectively, $\Phi$ is the dilaton,
$R_G$ is the scalar curvature
associated with the metric $G_{\mu\nu}$, and,
\ben \label{en2}
F^\ab_{\mu\nu} & = & \p_\mu A^\ab_\nu - \p_\nu A^\ab_\mu \, ,
\nonumber \\
H_{\mu\nu\rho} & = & (\p_\mu B_{\nu\rho} + 2 A^\ab_\mu L_{ab}
F^\bb_{\nu\rho}) + \hbox{~ cyclic permutations of $\mu$, $\nu$,
$\rho$}\, \nonumber \\
\een
are the field strengths associated with $A^\ab_\mu$ and $B_{\mu\nu}$.
$M$ is a $28\times 28$ matrix valued scalar field, satisfying,
\be \label{en3}
MLM^T = L, \qquad M^T=M\, ,
\ee
and,
\be \label{en4}
L = \pmatrix{ - I_{22} & 0 \cr 0 & I_{6} \cr }\, ,
\ee
where $I_n$ denotes $n\times n$ identity matrix.
The action \refb{en1} is invariant under an O(6,22) transformation:
\ben \label{elm1}
&& M\to \Omega M \Omega^T, \qquad A_\mu^\ab \to \Omega_{ab} A_\mu^\bb,
\nonumber \\
&& G_{\mu\nu}\to G_{\mu\nu}, \qquad \Phi\to \Phi, \qquad
B_{\mu\nu}\to B_{\mu\nu}\, ,
\een
where $\Omega$ is a $28\times 28$ matrix satisfying,
\be \label{elm2}
\Omega L \Omega^T = L\, .
\ee
This remains a valid symmetry of the action even after we include the
higher derivative terms in the action originating from the higher order
corrections in the string world-sheet theory. This fact
will be useful to us later.

We also define the canonical Einstein matric $g_{\mu\nu}$ as follows:
\be \label{en5}
g_{\mu\nu} = e^{-\Phi} G_{\mu\nu}\, .
\ee
In terms of $g_{\mu\nu}$, the action takes the form:
\ben \label{en6}
S & = & {1\over 32 \pi} \int d^4 x \sqrt{-g} \Big[
R_g - {1\over 2} g^{\mu\nu} \p_\mu \Phi \p_\nu \Phi - {1\over 12}
e^{-2\Phi}
g^{\mu\mu'} g^{\nu\nu'} g^{\rho\rho'} H_{\mu\nu\rho} H_{\mu'\nu'\rho'}
\nonumber \\
&& - e^{-\Phi}
g^{\mu\mu'} g^{\nu\nu'} F^\ab_{\mu\nu} (LML)_{ab} F^\bb_{\mu'\nu'}
+ {1\over 8} g^{\mu\nu} Tr(\p_\mu M L \p_\nu M L) \Big]\, .
\een

As in ref.\cite{SREV}, we use the normalization $\alpha'=16$. This
corresponds to a string world-sheet action of the form:
\be \label{en7}
{1\over 64 \pi} \int d^2 \xi G_{\mu\nu}(X) \p_\alpha X^\mu \p^\alpha
X^\nu + \cdots
\ee
where $\cdots$ denotes terms involving fermionic fields on the world
sheet, as well as the target space fields $B_{\mu\nu}$, $A^\ab_\mu$,
$\Phi$ and $M$.
We shall restrict ourselves to backgrounds characterized by the follwing
asymptotic forms of various fields:
\be \label{en8}
\langle g_{\mu\nu} \rangle = \eta_{\mu\nu}\, , \qquad
\langle e^{-\Phi} \rangle = {1\over g^2} , \qquad \langle M \rangle =
I_{28}\, , \qquad \langle B_{\mu\nu} \rangle = 0\, ,
\qquad \langle A^\ab_{\mu} \rangle = 0\, .
\ee
This gives
\be \label{en9}
\langle G_{\mu\nu} \rangle = g^2 \eta_{\mu\nu}\, .
\ee
{}From eqs.\refb{en6}, \refb{en7} and \refb{en9}
we see that the background
we are using corresponds to
the following values of the Newton's constant $G_N$ and
string tension $T$ measured by an asymptotic observer:
\be \label{e9}
G_N = 2\, , \qquad T = {g^2 \over 32 \pi}\, .
\ee

Most general electrically charged rotating black hole solutions in this
theory were constructed in ref.\cite{SBLACK}. (See also \cite{BGM,CVE}.)
We shall specialize on
the non-rotating extremal
black holes saturating the Bogomol'nyi bound. The most
general black hole solution of this type is given
by,\footnote{Ref.\cite{SBLACK} constructed the solution for $g=1$. The
solution for a general $g$ can easily be obtained from there by
appropriate rescaling of $e^{-\Phi}$ and $A^\ab_\mu$.}
\ben \label{e1}
ds^2 & \equiv & g_{\mu\nu} dx^\mu dx^\nu \nonumber \\
& = & - K^{-1/2} \rho dt^2 + K^{1/2} \rho^{-1} d\rho^2 + K^{1/2} \rho
(d\theta^2 + \sin^2\theta d\phi^2)\, ,
\een
\be \label{eqq2}
B_{\mu\nu}=0\, ,
\ee
\be \label{e6}
e^\Phi = K^{-1/2} \rho g^2 \, ,
\ee
\ben \label{e3}
A^\ab_t & = & -g {n^\ab\over \sqrt 2} {m_0\over K} \rho \sinh\alpha \, ,
\qquad \hbox{for} \qquad 1\le a \le 22\, , \nonumber \\
& = & -g {p^{(a-22)}\over \sqrt 2} {m_0 \over K} (\rho\cosh\alpha + m_0)
\, , \qquad \hbox{for} \qquad 23\le a \le 28\, , \nonumber \\
\een
\be \label{e4}
M = I_{28} + \pmatrix{ P nn^T & Q np^T \cr Q pn^T & Ppp^T\cr}\, ,
\ee
where $m_0$ and $\alpha$ are two real numbers,
$\vec n$ is a 22 dimensional
unit vector, $\vec p$ is a 6 dimensional unit vector, and,
\be \label{e2}
K = (\rho^2 + 2 m_0 \rho \cosh\alpha + m_0^2) \, ,
\ee
\ben \label{e5}
P & = & 2 {m_0^2\over K} \sinh^2\alpha \, , \nonumber \\
Q & = & - 2 {m_0 \sinh\alpha \over K} (\rho + m_0 \cosh \alpha)\, .
\een
The horizon and the singularity of this black hole coincide,
both being situated at $\rho=0$.

We define the electric charge $Q^\ab_{el}$ carried by the black
hole through the equations:
\be \label{e7}
F^\ab_{\rho t} \simeq {Q^\ab_{el} \over \rho^2} \qquad
\hbox{for large $\rho$} \, .
\ee
Eq.\refb{e3} then gives
\ben \label{e8}
Q^\ab_{el} & = & g {n^\ab \over \sqrt 2} m_0 \sinh\alpha\, , \qquad
\hbox{for} \qquad 1\le a \le 22\, , \nonumber \\
& = & g {p^{(a-22)}\over \sqrt 2} m_0 \cosh \alpha \, , \qquad
\hbox{for} \qquad 23 \le a \le 28 \, .
\een
Also, from eqs.\refb{e1}, \refb{e2} we see that
the ADM mass of the black hole is given by,
\be \label{e10}
m = {1\over G_N} {m_0\over 2} \cosh \alpha = {1\over 4} m_0 \cosh \alpha
\, .
\ee
It is customary to define the left and right components of the electric
charge vector as follows:
\ben \label{e12}
Q_R^\ab & = & 0 \qquad \hbox{for} \qquad 1\le a \le 22\, , \nonumber \\
& = & g {p^{(a-22)}\over \sqrt 2} m_0 \cosh \alpha \qquad \hbox{for}
\qquad 23 \le a \le 28\, , \nonumber \\
Q_L^\ab
& = & g {n^{(a)}\over \sqrt 2} m_0 \sinh \alpha \qquad \hbox{for}
\qquad 1 \le a \le 22\, , \nonumber \\
& = & 0 \qquad \hbox{for} \qquad 23\le a \le 28\, .
\een
{}From eqs.\refb{e10} and \refb{e12} we get,
\be \label{e11}
m^2 = {1\over 8g^2} \vec Q_R^2 \, ,
\ee
which is the standard Bogomol'nyi relation between mass and charge.
The independent parameters characterizing the black hole may be taken
to be $m$, $Q_L\equiv |\vec Q_L|$, $\vec n$ and $\vec p$. Eqs.\refb{e10},
\refb{e12} may now be inverted to give,
\be \label{eqq1}
m_0 = 4 \sqrt{m^2 - {\vec Q_L^2\over 8 g^2}}\, , \qquad
\alpha = \tanh^{-1}\Big( {Q_L\over 2\sqrt 2 gm}\Big)\, .
\ee

We shall now determine the position of the stretched horizon by examining
various fields close to the horizon.
For this we note that near the horizon ($\rho<<m_0$), the
string metric takes the form:
\ben \label{em1}
dS^2 & \equiv & G_{\mu\nu} dx^\mu dx^\nu = e^{\Phi} ds^2 \nonumber \\
& \simeq & -{\rho^2 \over m_0^2} g^2 dt^2 + g^2 d\rho^2 + g^2 \rho^2
(d\theta^2 + \sin^2\theta d\phi^2) \, .
\een
In the new coordinate system
\be \label{em2}
\brho = g \rho, \qquad \bt = t/m_0\, ,
\ee
the metric takes the form:
\be \label{em3}
dS^2 \simeq -\brho^2 d\bt^2 + d\brho^2 + \brho^2 (d\theta^2 + \sin^2\theta
d\phi^2) \, .
\ee
In the same coordinate system, other non-trivial fields near the horizon
are
\be \label{eqp1}
\p_{\brho} \Phi \simeq {1\over \brho}\, ,
\ee
\ben \label{eqp2}
F^\ab_{\brho\bt} & \simeq & -{n^\ab\over \sqrt 2} \sinh\alpha \qquad
\hbox{for} \quad 1\le a \le 22\, , \nonumber \\
& \simeq & {p^{(a-22)}\over \sqrt 2} \cosh\alpha \qquad
\hbox{for} \quad 23\le a \le 28\, .
\een
\be \label{eqp4}
M = I_{28} + \pmatrix{2 \sinh^2\alpha\, nn^T & -2 \sinh\alpha\cosh\alpha
\, np^T\cr -2 \sinh\alpha\cosh\alpha \, pn^T & 2 \sinh^2\alpha \,
pp^T\cr} \, .
\ee
Note that in this coordinate system, all dependence of the background
fields on the parameters $m_0$ and $g$ has disappeared, but
$F^\ab_{\brho\bt}$ and $M$ still depend on $\alpha$.
However, the $\alpha$ dependence of this background can be removed by
making an $O(6,22)$ transformation \refb{elm1} with the matrix
\be \label{elm3}
\Omega = \pmatrix{ \cosh\alpha \, nn^T & \sinh\alpha \, np^T \cr
\sinh\alpha \, pn^T & \cosh\alpha \, pp^T\cr}\, .
\ee
This gives,
\ben \label{elm4}
M' & \simeq & I_{28} \nonumber \\
F^{\prime \ab}_{\brho\bt} & \simeq & 0 \qquad
\hbox{for} \quad 1\le a \le 22\, , \nonumber \\
& \simeq & {p^{(a-22)}\over \sqrt 2} \qquad
\hbox{for} \quad 23\le a \le 28\, .
\een
Since \refb{elm1} represents a symmetry of the full effective action at
the string tree level, effects of world sheet quantum corrections, which
show up as higher derivative terms in the effective action, could be
studied as well in the transformed background represented by the primed
fields.
Since the background fields now do not depend on any parameter, we see
that the place where the target space field strengths become strong is
situated unambiguously at $\brho \sim 1$. This determines the location
of the stretched horizon to be at
\be \label{em4}
\brho =C\, ,
\ee
where $C$ is a pure number. This gives,
\be \label{em5}
\rho = C/g \equiv \eta\, .
\ee
It has been shown in the appendix that at this value of $\rho$, the local
Unruh temperature
is of the order of the Hagedorn temperature in string theory. Thus
the two definitions of stretched horizon coincide.

The area of the stretched horizon, calculated using the canonical metric
\refb{e1} is given by,
\be \label{e13}
A \simeq 4\pi \eta m_0 = 4\pi m_0 C/g\, .
\ee
Following the
Bekenstein-Hawking result, we define the entropy of the black hole
to be
\be \label{e14}
S_{B.H.} \equiv {A \over 4G_N} = {\pi \over 2} {m_0 C\over g}
= {2\pi C \over g} \sqrt{m^2 - {\vec Q_L^2\over 8g^2}}\, ,
\ee
where we have used eq.\refb{eqq1}.

Let us examine this equation in some detail. What kind of world sheet
corrections could produce such a modification of the entropy formula?
For this, we look at the
euclidean version of the string metric and the dilaton near the horizon:
\ben \label{ep0}
dS_E^2 & \simeq & \brho^2 d\btau^2 + d\brho^2 + \brho^2
(d\theta^2 + \sin^2\theta d\phi^2) \, , \nonumber \\
e^\Phi & \simeq & m_0^{-1} g \brho \, ,
\een
where $\btau=-i\bt$.
Let us define
\be \label{ekk1}
\bar\Phi = \Phi - \ln(g/m_0)\, , \qquad \bar g_{\mu\nu} = e^{-\bar
\Phi} G_{\mu\nu}\, ,
\ee
and introduce a new coordinate $\br$ through the relation
\be \label{ekk3}
\brho = \br^2/4\, .
\ee
Then eq.\refb{ep0} can be rewritten as
\ben \label{ekk2}
d\bar s_E^2 & \equiv & \bar g_{\mu\nu} dx^\mu dx^\nu \nonumber \\
& \simeq &
{1\over 4} \br^2 d\btau^2 + d\br^2 + {1\over 4} \br^2
(d\theta^2 + \sin^2\theta d\phi^2) \, , \nonumber \\
e^{\bar \Phi} & \simeq & \br^2/4 \, .
\een
If we take the coordinate $\btau$ to be periodic with period $4\pi$,
then the $\br,\btau$ part of the metric is non-singular at $\br=0$;
however the full metric is singular since the area of the transverse
sphere spanned by $\theta, \phi$ vanishes as
$\br\to 0$. Furthermore, the
dilaton also becomes singular at $\br=0$.
Let us suppose that the strong coupling effects on the
world sheet render the
solution finite at $\br=0$ by modifying the metric in such a way
that the area of the transverse sphere does not collapse at $\br=0$,
and the dilaton does not blow up as $\br\to 0$.\footnote{It
has been argued in ref.\cite{HOTS} that perturbative corrections on
the world-sheet does not modify the lowest order solution. However
this statement is valid only for a particular renormalization
scheme in the world sheet theory, and the metric that is relevant
for the entropy calculation may not be the one used in ref.\cite{HOTS}.
Put another way, if we insist on working with the metric of
ref.\cite{HOTS} then the surface term relevant for entropy calculation
may be modified by higher derivative terms in the effective action.}
This modification must
be such that it vanishes for $\br>>1$. A possible modification of
the solution near the point $\br=0$, that does this, is of the form:
\ben \label{ep1}
d\bar s_E^2 & \simeq & {1\over 4} \br^2 d\btau^2 + d\br^2 + f_1(\br)
(d\theta^2 + \sin^2\theta d\phi^2) \, , \nonumber \\
e^{\bar \Phi} & \simeq & f_2(\br) \, ,
\een
where $f_1$ and $f_2$ are two functions of $\br$ such that
$f_1(\br)\simeq\br^2/4$, $f_2(\br)\simeq\br^2/4$ for $\br>>1$,
and $f_1(0)$, $f_2(0)$ are finite and non-zero.
The important point to note is that $f_1$ and $f_2$ do not depend on
any of the parameters $m_0$, $\alpha$ or $g$.
This solution is non-singular if $\btau$ is taken to be a periodic
coordinate with period $4\pi$. The canonical metric $g_{\mu\nu}
=(m_0/g)\bar g_{\mu\nu}$ now takes the form:
\be \label{ep2}
ds_E^2 \simeq {m_0\over g }\big({1\over 4}\br^2
d\btau^2 + d\br^2 + f_1(\br)
(d\theta^2 + \sin^2\theta d\phi^2)\big)\, .
\ee
Near $\br=0$, the geometry of the space is that of a disc times a sphere
of radius
\be \label{ep3}
\sqrt{m_0 f_1(0) \over g }\, .
\ee
Then the entropy of the black hole, which is $(4G_N)^{-1}$ times the area
of this sphere, is given by,
\be \label{ep4}
{\pi m_0 f_1(0) \over 2 g }= {2 \pi f_1(0) \over g}
\sqrt{m^2 - {\vec Q_L^2\over 8g^2}}\, .
\ee
This has the same form as the expression for $S_{B.H.}$ given in
eq.\refb{e14}.

{\it
Although for this calculation we have postulated a specific scenario,
it should be clear from this discussion
that any modification of the surface terms\footnote{Besides the
extra surface terms that need to be added to the action\cite{GIBHAW},
the contribution from the bulk action is also given by a surface term
on shell, since $S= - \int d^4 x (\delta S/\delta \Phi(x))
= -\int d^4 x \p_\mu(\delta S/\delta (\p_\mu \Phi(x)))$.}
contributing to the entropy due to
the turning on of the stringy effects at $\brho\sim 1$,
will give an expression for the entropy of the form of \refb{e14}.
The main point is that the $\exp(-\Phi)$ term in front of the action
will produce a factor of $(m_0/g)$, and the rest of the surface terms,
which depend on the combination $(g/m_0)\exp(-\Phi)$,
the string metric and the other field strengths at
$\brho\sim 1$, will be totally independent of the parameters $m_0$,
$\alpha$ and $g$. Thus the contribution to the entropy will be
proportional to $m_0/g$, in agreement with eq.\refb{e14}.
This shows that if there is a modification of the Bekenstein-Hawking
formula for the black hole entropy due to string world-sheet effects,
then the correction is naturally of the form given in eq.\refb{e14}.}

We shall now compare the expression for the entropy given in eq.\refb{e14}
with the density of elementary string
states with the same mass and charge quantum numbers.
In the normalization convention we are using,
the mass formula for the Bogomol'nyi saturated
elementary string states is given by,
\be \label{e25}
m^2 = {\vec Q_R^2\over 8 g^2} =
{g^2 \over 8} \Big( {\vec Q_L^2 \over g^4} + 2 N_L -2 \Big)\, ,
\ee
where $N_L$ is the total oscillator contribution to the squared mass
from the left moving oscillators. There is no contribution to the mass
from the right moving oscillators, since in order to saturate the
Bogomol'nyi bound, the string state must be at the lowest level in
the right moving sector of the world-sheet. The degeneracy of such
states arises due to the many different ways the left moving oscillators
make up the total number $N_L$. This degeneracy has been calculated
many times (for a recent calculation, see \cite{RUSSO}) and is given
by,
\be \label{e27a}
d_{E.S.} \simeq \exp(4\pi \sqrt{N_L}) \, .
\ee
Thus, the entropy, calculated from the elementary string spectrum,
is given by,
\be \label{e27}
S_{E.S.} \equiv \ln d_{E.S.}
\simeq 4\pi \sqrt{N_L} \simeq {8 \pi \over g}
\sqrt{m^2 - {\vec Q_L^2 \over 8g^2}}\, .
\ee
This has the same dependence on $g$, $m$ and $\vec Q_L$ as $S_{B.H.}$
given in eq.\refb{e14}. The overall constant of proportionality
in the two expressions agree if we make the choice
\be \label{em7}
C=4 \, .
\ee
This shows that the modified Bekenstein-Hawking entropy of the black hole
does reproduce the density of elementary string states correctly.

\noindent{\bf Acknowledgement}: I wish to thank John Schwarz for several
illuminating discussions, without
which the detailed computation given here
would not have been performed. Some of these results were reported at
the workshop on `Physics at the Planck Scale, Puri, India,
Dec.12-21, 1994'. I would like to thank the organisers of
the workshop for hospitality.

\appendix

\renewcommand{\theequation}{\thesection.\arabic{equation}}

\sectiono{Stretched Horizon and Black Hole Thermodynamics}

In this appendix
we shall show that at the location of the stretched horizon
defined in eq.\refb{em5}, the local Unruh temperature does become
of the order of the Hagedorn temperature of string theory. This would
show that our definition of the stretched horizon
agrees with the definition of the stretched horizon advocated in
ref.\cite{SUSS}.  To do this we first euclideanize the canonical
metric \refb{e1},
and also choose an appropriate coordinate system in which the
$\rho,t$ part of the metric is non-singular. This is done through the
replacement
\be \label{e16}
t = i\tau, \qquad \rho = r^2\, .
\ee
The canonical metric near $\rho=0$ then takes the form
\be \label{e17}
ds_E^2 \simeq 4m_0 (dr^2 + {r^2\over 4 m_0^2} d\tau^2) + m_0 r^2
(d\theta^2 + \sin^2 \theta d\phi^2) \, .
\ee
The $r,\tau$ part of the metric describes a non-singular space, provided
$\tau$ describes a periodic coordinate with period $4\pi
m_0$:
\be \label{e18}
\tau \equiv \tau + 4\pi m_0\, .
\ee
{}From now on we shall make this identification.

The inverse of the
the local Unruh temperature is the proper period in the $\tau$ direction
at a fixed value of $\rho$ (or $r$). Using the metric
\refb{e17}, we see that at $\rho=\eta$ this is given by
\be \label{e19}
\beta_{Unruh}(\eta) \simeq 4 \pi \sqrt{m_0 \eta} = 4 \pi
\sqrt{m_0 C\over g} \, ,
\ee
where we have used the value of $\eta$ given in eq.\refb{em5}.

In order to calculate
the local Hagedorn temperature, we need to take into account the fact
that the string coupling, labelled by $e^\Phi$, is not a constant in
the black hole background, but actually varies with $\rho$. In fact,
at $\rho=\eta$ it is given by,
\be \label{e20}
e^{\Phi(\eta)} \simeq g^2 \eta/ m_0 = Cg/m_0 \, .
\ee
This gives
\be \label{e21}
\beta_{Hagedorn}(\eta) =
4\pi (2 + \sqrt 2) e^{-\Phi(\eta)/2}
= 4\pi (2 + \sqrt 2)  \sqrt{m_0\over C g}
\, .
\ee
Since $C\sim 1$
we see that at the stretched horizon, the local Unruh temperature
is indeed of the same order as the local Hagedorn
temperature.\footnote{Note
that since the string coupling vanishes as we approach the horizon, the
modification of the phase transition temperature due to the effects
discussed in ref.\cite{ATICKWITTEN} is expected to be small.}

\end{document}